\documentclass[12pt,english, smallheadings]{iopart}
\usepackage{iopams}
\newcommand{\ie}{{\rm i}}
\newcommand{\gam}{z}
\newcommand{\nn}{\nonumber \\}
\newcommand{\tO}[1]{O($#1$)}
\newcommand{\mO}[1]{{\rm O(}#1{\rm )}}
\newcommand{\twoeps}{(2\epsilon)}
\newcommand{\be}{\begin{equation}}
\newcommand{\ee}{\end{equation}}
\newcommand{\bea}{\begin{eqnarray}}
\newcommand{\eea}{\end{eqnarray}}
\newcommand{\citef}[2]{[#1]}
\begin{document}
\title[\tO{n}-symmetric Model of two $n$-Vector Fields]
{Critical Behavior of a General \tO{n}-symmetric Model of two
$n$-Vector Fields in $D=4-2\epsilon$}
\author{Yuri M. Pis'mak\dag, Alexej Weber\ddag, Franz J. Wegner\ddag}
\address{\dag\ Department of Theoretical Physics, State University
Saint-Petersburg, Russia}
\address{\ddag\ Institut f\"ur Theoretische Physik,
Ruprecht-Karls-Universit\"at Heidelberg, Germany}

\begin{abstract}
The critical behaviour of the \tO{n}-symmetric model with two
$n$-vector fields  is studied within the field-theoretical
renormalization group  approach in a $D=4-2\epsilon$ expansion.
Depending on the coupling constants the $\beta$-functions, fixed points
and critical exponents are calculated up to the one- and two-loop order, resp.
($\eta$ in two- and three-loop order).
Continuous lines of fixed points and \tO{n}$\times$\tO{2} invariant discrete
solutions were found. Apart from already known fixed points two new ones were
found. One agrees in one-loop order with a known fixed point, but differs from
it in two-loop order.
\end{abstract}


\submitto{\JPA}

\section{Introduction}

The renormalization group approach provides a natural framework
for  the understanding of critical properties of phase
transitions. A very large variety of critical phenomena can  be
described by so called $\phi^4$ models. The simple
\tO{n}-symmetric one-field model
\be\label{E-0a}
S_{\mO{n}}\left(\phi\right)=\frac{1}{2}\left[\left(\nabla\phi\right)^2
+ \tau\phi^2\right]+\frac{1}{4!}g \left(\phi^2\right)^2,
\ee
where $\phi=\left(\phi_1,\dots,\phi_n\right)$ is a real
$n$-component vector-field, while $\tau$ is a temperature-like
parameter and $g>0$, was extended in \cite{Fisher} to the
interplay of two vector-fields under the \tO{n}+\tO{m} symmetry
\bea\label{E-0b}
S_{\mO{n}+\mO{m}}\left(\phi_1,\phi_2\right)&=&
\frac{1}{2}\left[\left(\nabla\phi_1\right)^2+\left(\nabla\phi_2\right)^2
+ \tau_1\phi_1^2 +\tau_2\phi_2^2\right]
\nn
&+& \frac{1}{4!}\left[g_1(\phi_1^2)^2+g_2(\phi_2^2)^2
+g_3(\phi_1^2)(\phi_2^2)\right].
\eea
Six different fixed points were found. Three of them are always
unstable and the stability of three others depends on $n$ and $m$.
The \tO{n}+\tO{m} model has been used to describe
multicritical phenomena. We mention the critical behaviour  of
uniaxial antiferromagnets in a magnetic field parallel to the
field direction \cite{Fisher} and the SO(5)-theory of high-$T_\mathrm{c}$
superconductors \cite{Zhang, Aharony, Review}. Also interesting
phenomena of inverse symmetry breaking, symmetry nonrestoration
and reentrant phase transitions were reported \cite{Weinberg,
Pinto}. This model as well as model (\ref{E-0c}) presented below have also been investigated in \cite{Pelissetto}.

Recently frustrated spin systems with noncollinear or canted spin
ordering have been the object of intensive research
\cite{Vicary-299, Vicary-622, Vicary-623, Vicary-623b}.
Examples are helical magnets and layered triangular Heisenberg
antiferromagnets \cite{Kawamura-1988}. In the corresponding action
\bea \label{E-0c}
S_{\mO{n}\times\mO{2}}
\left(\phi_1,\phi_2\right)&=&
\frac{1}{2}\left[\left(\nabla\phi_1\right)^2+\left(\nabla\phi_2\right)^2
+ \tau\left(\phi_1^2 +\phi_2^2\right)\right]
\nn
&+& \frac{1}{4!}u\left(\phi_1^2+\phi_2^2\right)^2+\frac{1}{4!}v
\left[ (\phi_1 \phi_2)^2-(\phi_1^2)(\phi_2^2)\right]
\eea
the scalar product $\phi_1 \phi_2$ is present
\cite{Kawamura-1988-7, Kawamura-1988-10, Kawamura-1988-11}. Both
fields have $n$ components and the model possesses the \tO{n}$\times$\tO{2}
symmetry.  In the $4-2\epsilon$ expansion, the number of fix
points (FP) and their stability depend on $n$, however different
theoretical methods lead to contradictory results \cite{Review}.

The results based on 3-loop renormalization group calculations  \cite{Antonenko1, Antonenko, Loison} show that in
three-dimensional chiral magnets with $n = 2, 3$ critical fluctuations destroy
continuous phase transitions converting them into the first-order ones, i. e. the
chiral class of universality does not exist. On the other hand, the analysis of
the higher-order -- 5-loop and 6-loop -- $3D$ RG expansions reveals a new stable fixed
point for physical values of $n$ \cite{Pelissetto1}. This new fixed point turns out to be a focus \cite{Calabrese}
that governs the critical behavior of the system in a somewhat unusual way.
It was found to exist only for $n < 6$ \cite{Calabrese1} having no generic relation to
the stable chiral fixed point seen at small $\epsilon$ and large $n$. The situation
in two dimensions seems to be similar \cite{Calabrese2}.

The major part of the results obtained within other approaches ("exact"
renormalization group, Monte Carlo simulations, etc.) may be considered as favoring
the fluctuation-induced first-order chiral transitions for $n = 2, 3$ \citef{23-29}{Tissier1,Tissier2,Tissier3,Delamotte,Itakura,Peles,Bekhechi}. Such
transitions are characterized by effective critical exponents that are non-universal
and depend on the magnet or antiferromagnet studied. Arguments were presented \cite{Delamotte1}
that the new chiral fixed point found in \cite{Pelissetto1} may be an artefact produced by rather long RG expansions. For detailed discussion and most recent results see, e. g. \cite{Loison1, Thanh}.

The purpose of this paper is to investigate the critical behaviour
of the general \tO{n} symmetric theory
\bea \label{E-1}
S_{\mO{n}}\left(\phi_1,\phi_2\right)
&=S_0(\phi_1,\phi_2,\tau)+S_{\rm int}(\phi_1,\phi_2,g), \nn
S_0(\phi_1,\phi_2,\tau)&=
\frac 12\left(\sum_{k=1}^2(\nabla\phi_k)^2
+\sum_{k=1}^3\tau_k{\cal I}_k\right),
\\[1mm]
S_{\rm int}(\phi_1,\phi_2,g)&=\frac 18\sum_{k,l=1}^{3}
{\cal I}_k g_{kl} {\cal I}_l=\frac 18{\cal I}g{\cal I} \nonumber
\eea
of two classical  fields with $n$ components respectively,
with
\be
\left(\begin{array}c
{\cal I}_1 \\ {\cal I}_2 \\ {\cal I}_3
\end{array}\right) =
\left(\begin{array}c \phi_1^2 \\ \phi_2^2 \\ \sqrt 2 \phi_1\phi_2
\end{array}\right).
\ee
$g$ is assumed to be symmetric, $g_{ji}=g_{ij}$. Whenever possible we use only
$g_{ij}$ with $i\le j$.
The model  (\ref{E-1}) becomes \tO{n}+\tO{n} symmetric when
$\tau_3=g_{13}=g_{23}=g_{33}=0$. On the other hand, setting
\be
\tau_1=\tau_2, \quad \tau_3=g_{13}=g_{23}=0, \quad
g_{11}=g_{22}, \quad g_{12}= g_{11}-g_{33} \label{On2}
\ee
leads to the \tO{n}$\times$\tO{2} model of frustrated spins.

As a function of $n$ we find 10 FPs in total. To our knowledge the FPs we denote
by {\it RS 2.1b} and {\it RS 2.3} are new ones. The FP {\it RS 2.1b} is
remarkable, since in 1-loop order it coincides with the FP {\it RS 2.1a}, which
describes two decoupled isotropic systems. {\it RS 2.1b} shows in order
$\epsilon^{3/2}$ a coupling between both systems for general $n$.

In the next section we give the expression of the $\beta$-function and of
various anomalous dimensions, which allow the determination of the critical
exponents $\eta$, $\nu$ and $\omega$ and the cross-over exponents in one-loop
order for model (\ref{E-1}). In section \ref{Field_rotations}
we consider orthogonal transformations between the two
fields $\phi_1$ and $\phi_2$. As a consequence there will be discrete FPs
(invariant under this transformation) and lines of FPs.
Then we classify the solutions according to the behaviour in the large $n$
limit. In section \ref{finite_n} the various fixed points are determined and
the corresponding critical exponents are given for finite $n$.
If in some range of $n$ the FP becomes complex, we determine in order $\epsilon$
the limit $n_c$, where it becomes complex (for positive $n$ only).
Comparison is made with the known models (\ref{E-0a}-\ref{E-0c}) in section
\ref{subcases}. A summary concludes the paper.

\section{The $4-2\epsilon$ Expansion\label{eps_exp}}

The expression for the critical exponents can be taken from the review article
by Br\'ezin, le Guillou, and Zinn-Justin\cite{Domb-Green}. Writing
\be
S_{\rm int} = \frac 1{4!} g_{i\alpha,j\beta,k\kappa,l\lambda} \phi_{i\alpha}
\phi_{j\beta} \phi_{k\kappa} \phi_{l\lambda}
\ee
one obtains
\bea \fl
g_{i\alpha,j\beta,k\kappa,l\lambda} =&
\frac 1{\upsilon_{i,j} \upsilon_{k,l}} g_{\rho_{i,j},\rho_{k,l}}
\delta_{\alpha\beta} \delta_{\kappa,\lambda}
+\frac 1{\upsilon_{i,k} \upsilon_{j,l}}g_{\rho_{i,k},\rho_{j,l}}
\delta_{\alpha\kappa} \delta_{\beta\lambda}
+\frac 1{\upsilon_{i,l} \upsilon_{j,k}}g_{\rho_{i,l},\rho_{j,k}}
\delta_{\alpha\lambda} \delta_{\beta\kappa}, \\
& \upsilon_{1,1} = \upsilon_{2,2} = 1, \ \upsilon_{1,2}=\upsilon_{2,1} = \sqrt
2, \ \rho_{1,1} = 1, \rho_{2,2} = 2, \ \rho_{1,2}=\rho_{2,1} = 3.
\eea

The six $\beta$ functions $\beta_{ij}\equiv \mu \partial_\mu g_{ij}$,
where $\mu$ is an auxiliar parameter with the critical dimension
$1$, can be written in 1-loop order
\be \label{WV1}
\beta_{ij} = -2\epsilon g_{ij} + \frac 12 (n+8) g_{ik}g_{kl} +
\frac 12 C_{ij,kl,mn} g_{kl}g_{mn}
\ee
with
\be \label{beta_coeff} \fl
\begin{array}{cl}
i,j & C_{ij,kl,mn} g_{kl}g_{mn} \\
1,1 & -8g_{12}^2+2g_{12}g_{33}+g_{33}^2 \\
1,2 & -6g_{11}g_{12}-6g_{12}g_{22}-4g_{13}g_{23}
 +g_{11}g_{33}+4g_{12}^2+2g_{13}^2+g_{22}g_{33}+2g_{23}^2+g_{33}^2 \\
1,3 & -6g_{12}g_{23}-3g_{13}g_{33}+6g_{12}g_{13}+3g_{23}g_{33} \\
2,2 & -8g_{12}^2+2g_{12}g_{33}+g_{33}^2\\
2,3 & -6g_{12}g_{13}-3g_{23}g_{33}+6g_{12}g_{23}+3g_{13}g_{33}\\
3,3 & -2g_{13}^2 -2g_{23}^2 -6 g_{33}^2 +2g_{11}g_{33} +8g_{12}g_{33}
 +4g_{13}g_{23} +2g_{22}g_{33}
\end{array}
\ee
We have rescaled the couplings by a factor $8\pi^2$ as usual.

The FPs $g^*$ are the solutions of $\beta_{ij}(g^*)=0$. We observe that
(\ref{E-1}) is symmetric under the simultaneous interchange of
$g_{11}$ with $g_{22}$ and $g_{13}$ with $g_{23}$. The simultaneous change of
signs of $g_{13}$ and $g_{23}$ leaves the solution of
(\ref{WV1}) invariant.

The stability matrix
\be\label{the-4-eps-omega}
\omega_{ij,kl}=\partial\beta_{ij}(g)/\partial g_{kl}|_{g=g^*}
\ee
is easily obtained. The eigenvalues of (\ref{the-4-eps-omega}) are
the critical exponents $\omega$.

Similarly the critical exponents $\eta$ are obtained from the
eigenvalues $\gamma_\Phi^*$ of the symmetric $2\times 2$ matrix
$\gamma_\Phi$ at $g=g^*$,
\bea \label{the-4-eps-gamma-phi}
\fl \left\{\gamma_\Phi\right\}_{11} &=& \frac 1{16}\big(
2(n+2)g_{11}^2 +(n+2)g_{23}^2 +(n+1)g_{33}^2 +2ng_{12}^2
+4g_{12}g_{33} + 3(n+2)g_{13}^2\big), \nn
\fl \left\{\gamma_\Phi\right\}_{21}&=& \frac{\sqrt 2 (n+2)}{16}\big(
(g_{11}+g_{12}+g_{33})g_{13}+(g_{22}+g_{12}+g_{33})g_{23}
\big), \\
\fl \left\{\gamma_\Phi\right\}_{22}&=& \frac 1{16}\big(
(n+2)g_{13}^2 +2(n+2)g_{22}^2 +2ng_{12}^2 +3(n+2)g_{23}^2
+(n+1)g_{33}^2 +4g_{12}g_{33}\big), \nonumber
\eea
calculated at the specific FP, with respect to
$\eta_i=2\gamma_{\Phi_i}^*$.

The critical behaviour of perturbations bilinear in the fields $\phi$ are
governed by the expression for $1/\nu-2$ given by Br\'ezin et al, which
as function of the $n$ components of the fields can be written
\be
\left(\frac 1{\nu} -2 \right)_{i\alpha,j\beta;k\kappa,l\lambda} =
d^{(1)}_{i,j,k,l} \delta_{\alpha\beta} \delta_{\kappa\lambda}
+d^{(2)}_{i,j,k,l} \delta_{\alpha\kappa} \delta_{\beta\lambda}
+d^{(2)}_{i,j,l,k} \delta_{\alpha\lambda} \delta_{\beta\kappa}.
\label{dnu}
\ee

Eigenfunctions of this matrix are of three types:\\
(i) They may be \tO{n} symmetric corresponding to the variation of the
$\tau_i$. Thus one applies eigenfunctions of type $a_{kl}\delta_{\kappa\lambda}$
to (\ref{dnu}) and with
\be
\hat d^{(1)}_{\rho_{i,j},\rho_{k,l}} = \upsilon_{i,j} \upsilon_{k,l}
d^{(1)}_{i,j,k,l}, \
\hat d^{(2)}_{\rho_{i,j},\rho_{k,l}} = \upsilon_{i,j} \upsilon_{k,l}
(d^{(2)}_{i,j,k,l}+d^{(2)}_{i,j,l,k})
\ee
the eigenvalues are those of the $3\times 3$ matrix
\be
\gamma_{\tau} = n\hat d^{(1)} + \hat d^{(2)},
\ee
which in one-loop order reads
\be\label{the-4-eps-gamma-tau}
\gamma_{\tau}= -\frac 12 \left(
\begin{array}{ccc}
  (n+2)g_{11} & n g_{12}+g_{33} & (n+2)g_{13}  \\
  n g_{12}+g_{33} & (n+2)g_{22} & (n+2)g_{23}  \\
  (n+2)g_{13}  & (n+2)g_{23} & 2g_{12}+(n+1)g_{33}
\end{array}
\right)_{g=g^*}.
\ee
(ii) They may be of type $a_{k,l}b_{\kappa,\lambda}$ with $a$ and $b$ symmetric
in the indices, and $b_{\kappa,\kappa}=0$. They yield cross-over exponents which
are obtained
from the eigenvalues of the $3\times3$ matrix
\be \label{gamma_crs}
\gamma_{\rm cr,s} = \hat d^{(2)},
\ee
which in one-loop order reads
\be
\gamma_{\rm cr,s} = -\frac 12 \left(
\begin{array}{ccc}
 2g_{11} & g_{33} & 2g_{13} \\
 g_{33} & 2g_{22} & 2g_{23} \\
 2g_{13} & 2g_{23} & 2g_{12}+g_{33}
\end{array} \right)_{g=g^*}. \label{the-4-eps-gamma-cr}
\ee
(iii) Finally they may be of type $a_{k,l}b_{\kappa,\lambda}$, but now with both
$a$ and $b$ antisymmetric in their indices. They are obtained from
\be
\gamma_{\rm cr,a} = 2(d^{(2)}_{12,12}-d^{(2)}_{12,21}),
\ee
which in one-loop order reads
\be
\gamma_{\rm cr,a} = -g_{12}+\frac 12 g_{33}. \label{gamma_cra}
\ee

The various $\gamma$s given here are the anomalous dimensions in terms of the
length scale. The full dimension $y$ is written
\be
y_i = D-\frac{D-2}2 N \pm\gamma_i
\ee
for perturbations homogeneous in $\phi$ of order $N$.
For $\gamma^*_{\Phi}$ the minus sign
applies, whereas for the other exponents the plus sign has to be taken.
The first two contributions in the last
expression are the bare exponents valid for the trivial fixed point, whereas the
last term constitutes the anomalous contribution. If one
singles out a linear combination of the scalar products $\cal I$ as multiplied
by the temperature difference $\tau$ from the critical point, then the singular
part of the free energy shows the scaling behavior
\be
F_{\rm sing}(\tau,\{\mu_i\})
= |\tau|^{D\nu}F_{\rm sing,\pm}(\{\frac{\mu_i}{|\tau|^{\Delta_i}}\})
\label{Fsing}
\ee
near criticality, where $\tau$ and $\mu_i$ are multiplied by scaling operators.
$\nu$ obeys $y_{\tau}=1/\nu$ and the gap-exponents $\Delta_i$ are related to the
$y_i$ by
\be
\Delta_i = \frac{y_i}{y_{\tau}}=\nu y_i. \label{gapexp}
\ee
In the special case of operators bilinear in $\phi$ the exponents $\Delta_i$
are cross-over exponents.

\section{Field rotations\label{Field_rotations}}

One may perform a rotation between the fields $\phi_1$ and $\phi_2$ in the model
(\ref{E-1}),
\be\label{WV19}
\left(\begin{array}c \phi'_1 \\ \phi'_2 \end{array}\right)
=\left(\begin{array}{cc} \cos (\varphi) & \sin (\varphi) \\
-\sin (\varphi) & \cos (\varphi) \end{array}\right)
\left(\begin{array}c \phi_1 \\ \phi_2 \end{array}\right).
\ee

Performing the rotation (\ref{WV19}) yields
\be\label{We4}
\fl \left(\begin{array}c {\cal I}'_1 \\ {\cal I}'_2 \\ {\cal I}'_3
\end{array}\right)
= M \left(\begin{array}c {\cal I}_1 \\ {\cal I}_2 \\ {\cal I}_3
\end{array}\right), \quad
M=\left( \begin{array}{ccc}
\frac 12+\frac 12\cos(2\varphi) & \frac 12-\frac 12\cos(2\varphi) & \sqrt{\frac
12} \sin(2\varphi) \\
\frac 12-\frac 12\cos(2\varphi) & \frac 12+\frac 12\cos(2\varphi) & -\sqrt{\frac
12} \sin(2\varphi) \\
-\sqrt{\frac 12} \sin(2\varphi) & \sqrt{\frac 12} \sin(2\varphi) &
\cos(2\varphi)
\end{array} \right)
\ee
The  matrix $M$ is orthogonal
and the interaction transforms according to
\be
S_{\rm int}(\phi'_1,\phi'_2,g') = \frac 18 {\cal I'}^T g' {\cal I'}, \quad
g' = M g M^T.
\ee
Obviously both sets of couplings describe the same critical behavior.
One finds that
\be
a_1= g_{11}+g_{22}+2g_{12}, \quad
a_2= g_{11}+g_{22}+g_{33} \label{inv}
\ee
are invariant under the rotations, whereas
\bea
a_{31} = g_{11}-g_{22}, && \quad a_{32} = \sqrt 2 (g_{13}+g_{23}), \\
a_{41} = -g_{11}+2g_{12}-g_{22}+2g_{33}, && \quad
a_{42} = -\sqrt 8 (g_{13}-g_{23})
\eea
transform according to
\be \label{we2}
\left(\begin{array}c a'_{31} \\ a'_{32} \end{array}\right)
= \left(\begin{array}{cc} \cos(2\varphi) & \sin(2\varphi) \\
-\sin(2\varphi) & \cos(2\varphi) \end{array}\right)
\left(\begin{array}c a_{31} \\ a_{32} \end{array}\right)
\ee
and
\be \label{we3}
\left(\begin{array}c a'_{41} \\ a'_{42} \end{array}\right)
= \left(\begin{array}{cc} \cos(4\varphi) & \sin(4\varphi) \\
-\sin(4\varphi) & \cos(4\varphi) \end{array}\right)
\left(\begin{array}c a_{41} \\ a_{42} \end{array}\right).
\ee
For the interactions invariant under \tO{n}$\times$\tO{2} the amplitudes
$a_{31},\
a_{32},\ a_{41},\ a_{42}$ have to vanish. For all other interactions we may
choose $\varphi$. We will choose it so that
\be \label{we1}
a_{42}=0, \mbox{ that is } g_{23}=g_{13}.
\ee
In the following section we will derive
the FPs of (\ref{WV1}) with the condition (\ref{we1}), from which all
other fixed points can be obtained by means of the transformations
(\ref{we2},\ref{we3}) leaving the expressions (\ref{inv}) invariant.

\section{The Classification of the Fixed Points in the Large $n$ Limit}

\subsection{The form of the projectors}

In the large $n$ limit we may neglect the last term in (\ref{WV1}).
We express $g$ in terms of the matrix $p$,
\be
g=4\epsilon p /(n+8).
\ee
We see that at criticality ($\beta_{ij}\equiv 0$) and in the limit
$n\rightarrow\infty$ the matrix  $p$ becomes idempotent: $p=p^2$.
The only eigenvalues of idempotent matrices are 0 and 1. Thus depending on the
number $k$ of eigenvalues 1 there are four types of symmetric ($3\times 3$)
idempotent matrices $p^{(k)}$
\be \label{WV2}
p^{(0)}_{ij}=0, \ \ \  p^{(1)}_{ij}=\gam_i \gam_j, \ \ \
p^{(2)}_{ij}=\delta_{ij}-\gam_i \gam_j, \ \ \
p^{(3)}_{ij}=\delta_{ij}; \ \ \ i,j=1,2,3,
\ee
with the restriction
\be \label{WV3}
\gam_1^2+\gam_2^2+\gam_3^2=1.
\ee
Next the solution of (\ref{WV1}) in the limit $n\rightarrow\infty$ is calculated
by considering the first two orders  in $1/(n+8)$ to $g^*$.
This yields further conditions on $\gam$ for the classes $p^{(1,2)}$.

\subsection{The class $p^{(0)}$}

This class consists of the trivial FP $g^*=4\epsilon
p^{(0)}/(n+8)=0$ only. The stability matrix
\be \label{WV4}
\omega_{ij}= -\twoeps\delta_{ij}
\ee
is diagonal as we can see from (\ref{WV1}). All its eigenvalues
are negative and the FP is unstable. This FP is exact and  remains
invariant under the orthogonal transformations.

\subsection{The class $p^{(1)}$}

Here,  the ansatz
\be \label{WV5}
g_{ij}^*=
\frac{4\epsilon}{(n+8)}\gam_{i}\gam_{j}+\frac{4\epsilon}{(n+8)^{2}}h_{ij}+O\left(\frac{1}{(n+8)^{3}}\right)
\ee
with the symmetric matrix  $h$ is put into the $\beta$-functions
(\ref{WV1}). We neglect the terms of higher order in $1/(n+8)$ and obtain
\be \fl
-\frac{\epsilon^{2}\gam_{i}\gam_{j}}{n+8}-\frac{\epsilon^{2}h_{ij}}{(n+8)^{2}}
+\frac{\epsilon^{2}}{n+8}\left(\gam_{i}\gam_{k}+\frac{h_{ik}}{n+8}\right)\left(\gam_{k}\gam_{j}+\frac{h_{kj}}{n+8}\right)+\frac{\epsilon^{2}c_{ij}}{(n+8)^{2}}\cong
0,\label{WV6}
\ee
where
\be \label{WV7}
c_{ij}\equiv C_{ij,kl,mn}\gam_{k}\gam_{l}\gam_{m}\gam_{n}.
\ee
The equation for the terms of  first order in  $1/(n+8)$ gives the
already known condition (\ref{WV3}). From the equation for terms
of second order we obtain
\be \label{WV8}
-h_{ij}+\gam_{i}(\gam_{k}h_{kj})+\gam_{j}(\gam_{k}h_{ki})+c_{ij}=0.
\ee
Two of these six equations fix $\gam$, the remaining four can be
used to determine  $h$. We multiply  (\ref{WV8}) by  $\gam_{i}$
and sum over $i$
\be
-\gam_{i}h_{ij}+\gam_{k}h_{kj}+\gam_{j}(\gam_{k}h_{ki}\gam_{i})+\gam_{i}c_{ij}=0.
\ee
With $c_{j}\equiv\gam_{i}c_{ij}$ we obtain
\be
\gam_{j}(\gam_{k}h_{ki}\gam_{i})=-c_{j},
\ee
or
\be \label{WV9}
\gam_{i}c_{j}-\gam_{j}c_{i}=0.
\ee
The constants $c_{ij}=c_{ji}$ in (\ref{WV6}) can be calculated
with (\ref{WV1}) and (\ref{beta_coeff}).
The constants  $c_i=c_{ij}\gam_j$ then read
\bea
c_1 &=& \left(\gam _3^2-2 \gam_1 \gam_2\right) \left(7 \gam_1^2
\gam_2+3 \gam_2^3+4 \gam_2 \gam_3^2-2 \gam_1
\left(\gam_2^2+\gam_3^2\right)\right),
\\[3mm]
c_2 &=& \left(\gam_3^2-2 \gam_1 \gam_2\right) \left(\gam_1
\left(3 \gam_1^2-2 \gam_1 \gam_2+7 \gam_2^2\right)
+2 \left(2 \gam_1-\gam_2\right) \gam_3^2\right),
\\[3mm]
c_3 &=& 3 \gam_3 \left(2 \gam_1 \gam_2-\gam_3^2\right)
\left(\left(\gam_1-\gam_2\right)^2+2 \gam_3^2\right).
\eea
Two of the three equations (\ref{WV9}) turn out to be identical.
With (\ref{WV3}) we obtain the following  conditions on $\gam$:
\bea \label{WV10}
&& (1-\gam_{12}^2) (4-\gam_{12}^2)\gam_{12}
(\gam_1-\gam_2)=0,
\nn[3mm]
&& (1-\gam_{12}^2)(4-\gam_{12}^2)\gam_{12}\gam_3^2=0, \\[3mm]
&& \gam_{12}:=\gam_1+\gam_2. \nonumber
\eea
Thus solutions are given by
\be
\gam_{12}=0,\pm 1,\pm 2, \pm\sqrt 2,
\ee
where the first solutions can be read off immediately from the eqs.
(\ref{WV10}), whereas the last pair of solutions follows from
$\gam_1-\gam_2=0$, $\gam_3=0$ and eq. (\ref{WV3}). This last solution
describes an \tO{n}$\times$\tO{2}-invariant interaction. Due to the ansatz
(\ref{WV5}) a change of the sign of the $\gam$s does not alter the fixed point.
Thus $\gam_{12}$ and $-\gam_{12}$ yield the same class of fixed points. The
interaction can be
written
\be
S_{\rm int}^{(1)}=\frac{\epsilon}{2(n+8)} \big(\gam_i {\cal I}_i)^2
\ee
in this large $n$-limit. One realizes that the rotation (\ref{We4}) of $\cal I$
can be rewritten
\bea \label{rotI}
{\cal I}'_1+{\cal I}'_2 &=&  {\cal I}_1+{\cal I}_2, \\
\left(\begin{array}c {\cal I}'_1-{\cal I}'_2 \\ \sqrt 2 {\cal I}'_3
\end{array}\right) &=&
\left(\begin{array}{cc} \cos(2\varphi) & \sin(2\varphi) \\
-\sin(2\varphi) & \cos(2\varphi) \end{array}\right)
\left(\begin{array}c {\cal I}_1-{\cal I}_2 \\ \sqrt 2 {\cal I}_3
\end{array}\right)
\eea
Thus $\gam_{12}$ in
\be
\gam_i {\cal I}_i = \frac 12 \gam_{12}({\cal I}_1+{\cal I}_2)
+\frac 12 (\gam_1-\gam_2)({\cal I}_1-{\cal I}_2) + \gam_3{\cal I}_3
\ee
stays constant, whereas $\gam_1-\gam_2$ and $\gam_3$ vary under rotation
with
\be
(\gam_1-\gam_2)^2+2\gam_3^2 = 2-\gam_{12}^2. \label{rotII}
\ee
Thus for $\gam_{12}\ne\pm\sqrt 2$ one obtains a whole continuum of solutions.

The eigenvalues of the stability matrix are determined in appendix
\ref{stabmat}. In leading order they are independent of $\gam_{12}$.
Similarly one can determine the other exponents from
eqs. (\ref{the-4-eps-gamma-phi}-\ref{gamma_cra}) and obtain in the
limit of large $n$
\bea
\omega &=& \left\{\twoeps,0\ (2\times),-\twoeps\ (3\times)\right\}, \quad
\gamma_{\tau}^* = \left\{-\twoeps,0\ (2\times)\right\}, \nn
\gamma_{\rm cr}^* &=& \left\{
\frac{\twoeps}n(-1\pm\gam_{12}\sqrt{2-\gam_{12}^2}),
\frac{\twoeps}n(1-\gam_{12}^2)\ (2\times) \right\}, \\
\gamma^*_{\Phi} &=&
\left\{\frac{\twoeps^2}{8n}(1\pm\gam_{12}\sqrt{2-\gam_{12}^2})
\right\}. \nonumber
\eea
Here and in the following exponents appearing several times are indicated by
$(...\times)$. If a $\pm$
appears in an exponent, then exponents with both signs contribute. The exponent
$\gamma^*_{\rm cr,a}$ is always the last one of $\gamma^*_{\rm cr}$.

\subsection{The class $p^{(2)}$}

Here the ansatz
\be \label{WV11}
g_{ij}^*=
\frac{4\epsilon}{n+8}(\delta_{ij}-\gam_i \gam_j)+\frac{4\epsilon}{(n+8)^{2}}
h_{ij}+O\left(\frac{1}{(n+8)^{3}}\right),
\ee
with a symmetric matrix  $h$ is put into  (\ref{WV1}). This
leads to
\bea \label{WV12}
\fl && \frac{-\epsilon^2}{n+8}(\delta_{ij}-\gam_i\gam_j)-\frac{\epsilon^2
h_{ij}}{(n+8)^2}\nn
\fl &&
+\frac{\epsilon^2}{n+8}\left(\delta_{ik}-\gam_i\gam_k+\frac{h_{ik}}{n+8}\right)
\left(\delta_{kj}-\gam_k\gam_j+\frac{h_{kj}}{n+8}\right)+\frac{\epsilon^2
c_{ij}}{(n+8)^2} = O\left(\frac 1{(n+8)^3}\right),
\eea
with
\be
c_{ij} :=
C_{ij,kl,mn}(\delta_{kl}-\gam_k\gam_l)(\delta_{mn}-\gam_m\gam_n).
\ee

The equation for the first order terms in $1/(n+8)$ gives (\ref{WV3}) again.
The equation for the second order terms is
\be\label{WV13}
-h_{kj}\gam_i\gam_k+h_{ij}-h_{ik}\gam_k\gam_j+c_{ij}=0.
\ee
With the same arguments which led from (\ref{WV8}) to (\ref{WV10})
we now deduce conditions on $\gam$ corresponding to
(\ref{WV10}):
\bea \label{WV15}
&&(\gam_{12}^2+1)\gam_{12}^2(\gam_1-\gam_2) =0, \\[3mm]
&& (\gam_{12}^2+1) \gam_{12}\gam_3^2 =0.
\eea
Thus solutions are given by
\be
\gam_{12}=0,\pm\ie,\pm\sqrt 2,
\ee
where the first two solutions are immediately obvious from eqs. (\ref{WV15}) and
the last one follows from $\gam_1=\gam_2$, $\gam_3=0$, and
eq.(\ref{WV3}). This last solution represents an \tO{n}$\times$ \tO{2}-invariant
model.
The interaction can be written
\be
S_{\rm int}^{(2)} = \frac{\epsilon}{2(n+8)}
\left( {\cal I}_i{\cal I}_i-(\gam_i{\cal I}_i)^2 \right)
\ee
in the large $n$-limit. Note that ${\cal I}_i{\cal I}_i$ is invariant under
rotations (\ref{We4}). Thus the same argument concerning the invariance of
$\gam_{12}$ under rotations as for $p^{(1)}$ applies here. Again for
$\gam_{12}\ne\pm\sqrt 2$ one obtains a continuous set of models related by
the transformation (\ref{rotI} to \ref{rotII}).

The stability matrix $\omega$ yields in this limit eigenvalues opposite in sign
to those of $p^{(1)}$ (appendix \ref{stabmat}).
Similarly one determines the other exponents from eqs.
(\ref{the-4-eps-gamma-phi}-\ref{gamma_cra}) and obtains in the limit
of large $n$
\bea
\omega &=& \left\{\twoeps\ (3\times),0\ (2\times),-\twoeps\right\}, \quad
\gamma_{\tau}^* = \left\{-\twoeps\ (2\times),0\right\}, \nn
\gamma_{\rm cr}^* &=& \left\{\frac{\twoeps}n(-2+\gam_{12}^2),
\frac{\twoeps}n(-1\pm\sqrt{1+2\gam_{12}^2-\gam_{12}^4}),
\frac{\twoeps}n\gam_{12}^2 \right\}, \label{change1} \\
\gamma_{\Phi}^* &=&
\left\{\frac{\twoeps^2}{8n}(2\pm\gam_{12}\sqrt{2-\gam_{12}^2}),
\right\}. \nonumber
\eea

\subsection{The class $p^{(3)}$}

In the large $n$ limit one obtains $g^*=4\epsilon p^{(3)}/(n+8)$, which
yields the exponents in leading order
\bea
\omega &=&
\left\{\twoeps\ (6\times)\right\},
\quad
\gamma_{\tau}^* = \left\{-\twoeps\ (3\times)\right\}, \nn[-2mm]
\\[-2mm]
\gamma_{\rm cr}^* &=& \left\{\frac{-3\twoeps}n,
\frac{-\twoeps}n\ (2\times),
\frac{\twoeps}n \right\}, \quad
\gamma_{\Phi}^* = \left\{\frac{3\twoeps^2}{8n}\ (2\times)\right\}.
\nonumber
\eea

\section{Solutions for Finite $n$\label{finite_n}}

\subsection{Fixed Points}

In order to solve the eqs. (\ref{WV1}) for the couplings $g^*$ for finite $n$,
we observe that the 'gauge' condition $a_{42}=g_{13}-g_{23}=0$ yields
\bea
\beta_{11}-\beta_{22}
&=& -\frac 12(g_{11}-g_{22})(4\epsilon-(n+8)(g_{11}+g_{22})) = 0, \\
\beta_{13}-\beta_{23} &=& \frac{n+8}2(g_{11}-g_{22})g_{13} = 0, \\
\beta_{13} &=& -\frac 12 g_{13} (4\epsilon-(n+8)(g_{11}+g_{12}+g_{33})) =
0.
\eea
Thus we have to solve any of the two equations
\bea
g_{13}=0 && \quad g_{11}=g_{22} \label{sols1} \\
g_{13}=0 && \quad 4\epsilon-(n+8)(g_{11}+g_{22}) = 0 \label{sols2} \\
g_{11}=g_{22} && \quad 4\epsilon-(n+8)(g_{11}+g_{12}+g_{33})=0
\label{sols3}
\eea
together with the three equations
\bea
&\beta_{11}=\beta_{12}=0, \label{sols4} \\
&\beta_{33}=0. \label{sols5}
\eea
If $g_{13}=0$, then $\beta_{33}$ factors
\be
\beta_{33} = -\frac 12 g_{33}
(4\epsilon-(n+2)g_{33}+2g_{11}+2g_{22}+8g_{12}).
\ee
Then we distinguish the two cases
\bea
&& g_{33}=0, \label{sols5a} \\
&& 4\epsilon-(n+2)g_{33}+2g_{11}+2g_{22}+8g_{12}=0. \label{sols5b}
\eea
One obtains the following solutions from (\ref{sols1},\ref{sols4},\ref{sols5a})
\bea
\fl && g_{11}=g_{22}, \quad g_{13}=g_{23}=g_{33}=0 \nn
\fl && \begin{array}{lcccccc}
RS & g_{11}/\epsilon & g_{12}/\epsilon & a_1/\epsilon &
a_2/\epsilon & a_{41}/\epsilon & \gam_{12} \\
0.1 & 0 & 0 & 0 & 0 & 0 & -\\
2.1 & \frac 4{n+8} & 0 & \frac 8{n+8} & \frac 8{n+8} & -\frac 8{n+8} & 0 \\
1.3 & \frac 2{n+4} & \frac 2{n+4} & \frac 8{n+4} & \frac 4{n+4} & 0 & \pm\sqrt 2
\\
1.2 & \frac{2n}{n^2+8} & \frac{8-2n}{n^2+8} & \frac{16}{n^2+8} &
\frac{4n}{n^2+8} & -\frac{8(n-2)}{n^2+8} & 0
\end{array} \label{rep1}
\eea

Eqs. (\ref{sols1},\ref{sols4},\ref{sols5b}) yield for $g_{33}\not=0$
\bea
\fl && g_{11}=g_{22}, \quad g_{13}=g_{23}=0 \nn
\fl && \begin{array}{lccc}
RS & g_{11}/\epsilon & g_{12}/\epsilon & g_{33}/\epsilon
\\
2.1 & \frac 2{n+8} & \frac 2{n+8} & \frac 4{n+8} \\
1.2 & \frac 4{n^2+8} & \frac 4{n^2+8} & \frac{4(n-2)}{n^2+8} \\
3.1 & \frac{3n^2-2n+24+(n-6)\sqrt{n^2-24n+48}}{n^3+4n^2-24n+144} &
\frac{-n^2-6n+72+(n+6)\sqrt{n^2-24n+48}}{n^3+4n^2-24n+144} &
\frac{4(n^2+n-12-3\sqrt{n^2-24n+48})}{n^3+4n^2-24n+144} \\
2.2 & \frac{3n^2-2n+24-(n-6)\sqrt{n^2-24n+48}}{n^3+4n^2-24n+144} &
\frac{-n^2-6n+72-(n+6)\sqrt{n^2-24n+48}}{n^3+4n^2-24n+144} &
\frac{4(n^2+n-12+3\sqrt{n^2-24n+48})}{n^3+4n^2-24n+144}
\end{array} \label{rep2} \\
\fl && \begin{array}{lcccc}
RS & a_1/\epsilon & a_2/\epsilon & a_{41}/\epsilon & \gam_{12} \\
2.1 & \frac 8{n+8} & \frac 8{n+8} & \frac 8{n+8} & 0 \\
1.2 & \frac{16}{n^2+8} & \frac{4n}{n^2+8} & \frac{8(n-2)}{n^2+8} & 0 \\
3.1 & \frac{4(n^2-4n+48+n\sqrt{n^2-24n+48})}{n^3+4n^2-24n+144} &
\frac{2(5n^2+(n-12)\sqrt{n^2-24n+48})}{n^3+4n^2-24n+144} & 0 & - \\
2.2 & \frac{4(n^2-4n+48-n\sqrt{n^2-24n+48})}{n^3+4n^2-24n+144} &
\frac{2(5n^2-(n-12)\sqrt{n^2-24n+48})}{n^3+4n^2-24n+144} & 0 & \pm\sqrt 2
\end{array} \nonumber
\eea

Eqs. (\ref{sols2},\ref{sols4},\ref{sols5}) with $g_{11}\ne g_{22}$ yield
\bea
\fl && g_{13}=g_{23}=0, \nn
\fl && \begin{array}{lccc}
RS & g_{11,22}/\epsilon-\frac 2{n+8} & g_{12}/\epsilon &
g_{33}/\epsilon \\
1.1 & \pm\frac 2{n+8} & 0 & 0 \\
1.4 & \pm\sqrt{\frac{32(1-n)}{(n+8)^3}} & \frac 6{n+8} & 0 \\
2.3 & \pm\sqrt{\frac{-4(3n+22)(n-2)(n+2)(n+4)(n+14)} {(n+8)^3(n^2+4n+20)^2}} &
\frac{4(n+6)(n+4)}{(n+8)(n^2+4n+20)} & \frac{4(n^2-36)}{(n+8)(n^2+4n+20)}
\end{array} \label{rep3} \\
\fl && \begin{array}{lcccc}
RS & a_1/\epsilon & a_2/\epsilon & a_{41}/\epsilon & \gam_{12}\\
1.1 & \frac 4{n+8} & \frac 4{n+8} & -\frac 4{n+8} & \pm 1 \\
1.4 & \frac{16}{n+8} & \frac 4{n+8} & \frac 8{n+8} & \pm 2 \\
2.3 & \frac{4(3n^2+24n+68)}{(n+8)(n^2+4n+20)} &
\frac{8(n+4)(n-2)}{(n+8)(n^2+4n+20)} &
\frac{4(3n^2+16n-44)}{(n+8)(n^2+4n+20)} & \pm \ie
\end{array} \nonumber
\eea

Eqs. (\ref{sols3},\ref{sols4},\ref{sols5}) yield for $g_{13}\ne 0$
\bea
\fl && g_{11}=g_{22}, \quad g_{13}=g_{23} \nn
\fl && \begin{array}{lcccc}
RS & g_{11}/\epsilon & g_{12}/\epsilon & g_{33}/\epsilon
& (g_{13}/\epsilon)^2 \\
1.1 & \frac 1{n+8} & \frac 1{n+8}& \frac 2{n+8} & \frac 2{(n+8)^2} \\
1.4 & \frac 4{n+8} & \frac 4{n+8} & -\frac 4{n+8} & \frac{16(1-n)}{(n+8)^3} \\
2.3 & \frac{5n^2+24n-4}{(n+8)(n^2+4n+20)} &
\frac{(n+10)(n+14)}{(n+8)(n^2+4n+20)} &
\frac{-2(n+2)(n+14)}{(n+8)(n^2+4n+20)} &
\frac{-2(3n+22)(n-2)(n+2)(n+4)(n+14)}{(n+8)^3(n^2+4n+20)^2}
\end{array} \label{rep4} \\
\fl && \begin{array}{lcccc}
RS & a_1/\epsilon & a_2/\epsilon & a_{41}/\epsilon & \gam_{12} \\
1.1 & \frac 4{n+8} & \frac 4{n+8} & \frac 4{n+8} & \pm 1\\
1.4 & \frac{16}{n+8} & \frac 4{n+8}& -\frac 8{n+8} & \pm 2 \\
2.3 & \frac{4(3n^2+24n+68)}{(n+8)(n^2+4n+20)} & \frac{8(n-2)(n+4)}{(n+8)
(n^2+4n+20)} & -\frac{4(3n^2+16n-44)}{(n+8)(n^2+4n+20)} & \pm \ie
\end{array} \nonumber
\eea
We consider the solutions (\ref{rep1} to \ref{rep4}) as representative
solutions. They are denoted by {\it RS k.m}, where $k$ indicates that they
belong to $p^{(k)}$ in the large $n$ limit, and $m$ numbers the various
solutions.

There are three types of solutions:\\
(i) The solutions, which are invariant under \tO{n}$\times$\tO{2}. There is one
solution for each $k$, {\it RS 0.1, 1.3, 2.2,} and {\it 3.1},\\
(ii) solutions for which $a_{31}=a_{32}=0$, {\it RS 1.2, 2.1}, and\\
(iii) solutions for which $a$s can be different from 0, {\it RS 1.1, 1.4,} and
{\it 2.3}. The solutions can be seen both in (\ref{rep3}) and (\ref{rep4}). They
are
obtained from one another by a rotation by $\varphi=\pi/4$.\\
All solutions with the exception of the trivial fixed point {\it RS 0.1} have an
exponent $\omega=2\epsilon$ independent of $n$ in one-loop order, since
$\beta_{ij}=-2\epsilon g_{ij} +$ term bilinear in the $g$s and thus $\partial
\beta_{ij}/\partial g_{kl}|_{g=g^*} g^*_{kl} = 2\epsilon g^*_{ij}$.

For the solutions (i) of symmetry \tO{n}$\times$\tO{2} eq. (\ref{On2}) holds.
Then eqs. (\ref{the-4-eps-gamma-phi}, \ref{the-4-eps-gamma-tau},
\ref{the-4-eps-gamma-cr}) yield the eigenvalues
\bea
\gamma^*_{\Phi} = \left\{\frac{n+1}4 g_{11}^{*2} +\frac{3(n-1)}{16} g_{33}^{*2}
-\frac{n-1}4 g^*_{11}g^*_{33}\ (2\times)\right\}, \nn
\gamma^*_{\tau} = \left\{-(n+1)g^*_{11} +\frac{n-1}2 g^*_{33},
-g^*_{11}-\frac{n-1}2 g^*_{33}\ (2\times)\right\}, \label{gamo2} \\
\gamma^*_{\rm cr,s} = \left\{-g^*_{11}-\frac 12 g^*_{33},
-g^*_{11}+\frac 12 g^*_{33}\ (2\times)\right\}. \nonumber
\eea
All three sets of exponents contain two degenerate exponents. The first
exponent $\gamma^*_{\tau}$ yields $\nu$, the two other ones belong to
perturbations of type $\phi_1^2-\phi_2^2$ and $\phi_1\phi_2$. Thus
they yield cross-over exponents. The first cross-over exponent
$\gamma^*_{\rm cr}$ belongs to operators $b_{\kappa\lambda}
(\phi_{1\kappa}\phi_{1\lambda}+\phi_{2\kappa}\phi_{2\lambda})$,
the two equal exponents to $b_{\kappa\lambda}
(\phi_{1\kappa}\phi_{1\lambda}-\phi_{2\kappa}\phi_{2\lambda})$ and
$b_{\kappa\lambda} \phi_{1\kappa}\phi_{2\lambda}$ with symmetric
$b_{\kappa\lambda}$. The degeneracies are due to the \tO{2} invariance.

All other solutions to type (ii) and (iii) can be obtained by means
of field rotations as described in section \ref{Field_rotations}.
These solutions yield one exponent $\omega=0$ since the field rotations create
lines of fixed points. This exponent in not a true scaling exponent, but a
redundant one, since the perturbation is obtained from an infinitesimal
rotation between $\phi_1$ and $\phi_2$.

\subsection{Critical Exponents}

In the following we give the critical exponents of the various fixed points.

\paragraph{RS 0.1} This is the trivial (interaction free) fixed point. All
anomalous exponents $\gamma^*$ vanish
\be
\gamma_{\Phi}^*=\left\{0\ (2\times) \right\} \ \
\gamma_{\tau}^*=\left\{0\ (3\times) \right\} \ \
\gamma_{\rm cr}^*=\left\{0\ (4\times) \right\} \ \
\omega=\left\{-\twoeps\ (6\times) \right\}
\ee

\paragraph{RS 1.1} Representatives of these solutions are given in (\ref{rep3})
and (\ref{rep4}).
The critical exponents are given by
\bea
\fl \gamma_{\tau}^* &=&\left\{ -\frac{(n+2)\twoeps}{n+8},0\ (2\times)\right\},
\quad
\gamma_{\rm cr}^*= \left\{ -\frac{2\twoeps}{n+8},0\ (3\times) \right\},
\label{rl-p1-1} \\
\fl \gamma_{\Phi}^* &=& \left\{ \frac{(n+2)\twoeps^2}{4(n+8)^2},0
\right\}, \quad
\omega=\left\{\twoeps,-\twoeps\ (2\times)
,-\frac{(n+6)\twoeps}{n+8},-\frac{6\twoeps}{n+8},0\right\}.
\nonumber
\eea

\paragraph{RS 1.2} Representatives are given in (\ref{rep1}) and (\ref{rep2}).
The critical exponents are
\bea
\fl \gamma_{\Phi}^* &=& \left\{\frac{n (n^2-3n+8) \twoeps^2} {8(n^2+8)^2}
(2\times)\right\}, \quad
\gamma_\tau^*=\left\{-\frac{3 n
\twoeps}{n^2+8},\frac{(1-n) n \twoeps}{n^2+8},
\frac{(n-4)\twoeps}{n^2+8}\right\}, \nn
\fl \gamma_{\rm cr}^* &=& \left\{
-\frac{n\twoeps}{n^2+8}\ (2\times),
\frac{(n-4)\twoeps}{n^2+8} (2\times)
\right\}, \label{rl-p1-4} \\
\fl \omega &=&\left\{0,\twoeps,\frac{8(n-1) \twoeps}{n^2+8},
\frac{(4-n)(2+n)\twoeps}{n^2+8}, \right. \nn
\fl && \left. \frac{(4-n)(n-2)\twoeps}{n^2+8},
\frac{(2-n) (4+n) \twoeps}{n^2+8}\right\}. \nonumber
\eea
In the representation (\ref{rep1}) $g_{13}=g_{23}=0$ holds and the
$\gamma_\tau$ matrix
(\ref{the-4-eps-gamma-tau}) becomes a block-matrix and has the
eigenvalues
$\gamma_{\tau}^*=\{\gamma_{\tau_1}^*,\gamma_{\tau_2}^*,\gamma_{\tau_3}^*\}$
which  in the case of $g_{11}^*=g_{22}^*$ belong to the eigenvectors
$(1,1,0)$, $(1,-1,0)$ and $(0,0,1)$ respectively in our
convention. Thus the first entry represents an ordinary critical
exponent when $\tau_1=\tau_2$, the third entry is the critical
exponent of $\tau_3$, and the second entry as well as the exponents
$\gamma_{\rm cr}^*$ are related to the crossover.

\paragraph{RS 1.3} This solution is not only invariant under
\tO{n}$\times$\tO{2}, but even under \tO{2n}. $g^*$ is given in (\ref{rep1}).
Its critical exponents are
\bea \label{RS-1-3-FP-exp}
\fl \gamma_{\Phi}^* &=&\left\{\frac{(2n+2)\twoeps^2}{4(2n+8)^2} \
(2\times)\right\},
\quad \gamma_\tau^*=\left\{-2\frac{(2n+2)\twoeps}{2n+8},
-\frac{2\twoeps}{2n+8} \ (2\times)\right\},
\\
\fl \gamma_{\rm cr}^* &=& \left\{-\frac{2\twoeps}{2n+8}\ (4\times)\right\},
\ \ \omega=\left\{\twoeps,\frac{8\twoeps}{2n+8} \ (2\times),
\frac{(4-2n)\twoeps}{2n+8} \ (3\times) \right\}. \nonumber
\eea
The last two exponents of $\gamma^*_{\tau}$ belong to cross-over
exponents (discussion after (\ref{gamo2})). These exponents and all exponents $\gamma^*_{\rm cr}$ are equal.

\paragraph{RS 1.4} Its representative couplings are given in (\ref{rep3}) and
(\ref{rep4}). In one loop order one obtains the exponents
\bea
\fl \gamma_\Phi^* &=& \left\{\frac{(n^2+37n+16)\twoeps^2}{8(n+8)^3}
\pm (n+2)\frac{\sqrt{2(1-n)}\twoeps^2}{2(n+8)^{5/2}}\right\},
\nn
\fl \gamma_\tau^* &=& \left\{-\frac{(2+n)\twoeps}{2(n+8)}
\pm\frac{\sqrt{n^3+48n^2+32}\twoeps}{2(n+8)^{3/2}},
-\frac{3\twoeps}{n+8}\right\}, \label{rs14} \\
\fl \gamma_{\rm cr}^*
&=& \left\{-\frac{\twoeps}{n+8}\pm\frac{2\sqrt{2(1-n)}\twoeps}{(n+8)^{3/2}},
-\frac{3\twoeps}{n+8} (2\times)\right\}, \nn
\fl \omega &=& \left\{0,\twoeps,\frac{(6-n) \twoeps}{n+8},
\frac{(10-n)\twoeps}{n+8},
-\frac{(n+2)\twoeps}{2(n+8)}\pm\frac{\sqrt{n^2-188n+196} \twoeps }{2(n+8)}
\right\}. \nonumber
\eea

We consider the coupling in two loop order, since it yields in order $\epsilon$
the region in which the couplings are real. Using the representation
(\ref{rep3}) the couplings may be written
\bea
g^*_{11,22} &=& u^{(1)}\epsilon + u^{(2)}\epsilon^2 \pm V + O(\epsilon^3),
\label{U1} \\
V &=& v^{(1)}\sqrt w \epsilon + \frac{v^{(2)}}{\sqrt w} \epsilon^2, \label{V1}
\eea
with
\bea
u^{(1)} &=& \frac 2{n+8}, \nn
u^{(2)} &=& -\frac{2(n^3+24n^2-27n-160)}{(n+8)^4}, \nn
v^{(1)} &=& \frac 4{(n+8)^2}, \\
v^{(2)} &=& \frac{n^4+80n^3-2004n^2-880n+2560}{2(n+8)^4}, \nn
w &=& 2(1-n)(8+n), \nn
g^*_{12} &=& \frac 6{n+8}\epsilon + \frac{-n^3-66n^2+450n+832}{(n+8)^4}
\epsilon^2, \nn
g^*_{13} &=& g^*_{23} = g^*_{33} = 0. \nonumber
\eea
Now $V$ can be rewritten
\be
V=v^{(1)}\epsilon \sqrt{w+\frac{2v^{(2)}}{v^{(1)}}\epsilon}. \label{V2}
\ee
Thus with $w(n_0)=0$ the limit of real couplings is given by
\be
n_c=n_0-\frac{2v^{(2)}(n_0)}{v^{(1)}(n_0)w'(n_0)}\epsilon, \label{V3}
\ee
which in our case yields $n_c=1-\twoeps/48+O\twoeps^2$.

\paragraph{RS 2.1} Representatives in one loop-order are given in (\ref{rep1})
and (\ref{rep2}). Two of the exponents $\omega$ equal 0 for any $n$ in one-loop
order. One is due to the invariance under rotations between the fields $\phi$.
The other one indicates that there may branch off a second line of FPs. Indeed
one finds besides the FP of two decoupled systems $g^*_{11} = g^*_{22}$,
$g^*_{12} = g^*_{33} = g^*_{13} = g^*_{23} = 0$ (which we denote {\it RS 2.1a})
another solution with
$g^*_{11} = g^*_{22}$, $g^*_{12},g^*_{33} = O(\epsilon^2)$,
$ g^*_{13} = g^*_{23} = O(\epsilon^{3/2})$, which we denote {\it RS 2.1b}. Both
types of FPs agree in one-loop order, but differ in the next order. Note that
the first FP has $a_{31}=a_{32}=0$, whereas the second does not show this
symmetry. In the following we give the FPs and critical exponents in two-loop
order (for $\gamma^*_{\Phi}$ in three-loop order).

First the general scheme to obtain the FPs beyond first order is explained.
Let the $\beta$-function up to two-loop order read
\be
\beta_i = -\twoeps g_i + \sum_{pq} k_{ipq}g_pg_q + \sum_{pqr} l_{ipgr}g_pg_qg_r
+ ...,
\ee
where the indices $_i$, $_p$, $_q$, $_r$ replace the double indices $_{ij}$ and
expand the contributions in one-loop order
\be
k_{ipq} = k^0_{ipq} + \epsilon k^1_{ipq} + ...
\ee
and similarly the higher-loop orders. With
\be
g_i=\epsilon g_{1,i}+\epsilon^2 g_{2,i} + ... \label{gexp}
\ee
one obtains from $\beta_i(g^*)$ order $\epsilon^r$, $r>2$ the equation
\bea
B_{ij} g^*_{r-1,j} &=& {\rm r.h.s} \label{Beq} \\
B_{ij} &:=& -2\delta_{ij} + 2\sum_p k^0_{ipq} g^*_{1,p},
\eea
where the r.h.s. of the equation (\ref{Beq}) contains only $g^*_{r',q}$ with
$r'<r-1$.
The matrix $B$ is the matrix $\omega$ in one-loop order. If none of the
eigenvalues of this matrix vanishes, then eq. (\ref{Beq}) can be used to
calculate $g^*_r$ in increasing order $r$. If due to the rotation invariance
one of the eigenvalues vanishes, then the condition (\ref{we1}) reduces the
number of independent couplings by 1 and eliminates the vanishing eigenvalue.
If, however, a second eigenvalue vanishes, then the calculation has to be
modified. For this {\it RS 2.1} we assume $g_{22}=g_{11}$ and expand $g_{11}$,
$g_{12}$, $g_{33}$ as in eq. (\ref{gexp}),
but denote $g_u=g_{13}=g_{23}$  and expand
\be
g_u=\epsilon^{3/2} g_{1,u} + \epsilon^{5/2} g_{2,u} + ...
\ee
From now on the indices $i,p,q,...$ stand only for the double indices
$11,12,33$, but not for $13$.

Order $\epsilon^2$ of $\beta_i(g^*)=0$ is fulfilled by the
solutions of {\it RS 1.2}
\be
g^*_{1,11}=\frac 4{n+8}, \quad g^*_{1,12}=g^*_{1,33} = 0.
\ee
Order $\epsilon^{5/2}$ of $\beta_u=0$ yields
\be
B_{uu} g^*_{1,u} = 0.
\ee
Since for the FP {\it RS 2.1} $B_{uu}=0$, this is automatically fulfilled.
Next $\beta_i(g^*) = 0$ in order $\epsilon^3$ yields
\be
\sum_q B_{iq} g^*_{2,q} + k^0_{iuu} g^{*2}_{1,u}
+\sum_{pq} k^1_{ipq} g^*_{1,p}g^*_{1,q} + \sum_{pqr} l_{ipqr}
g^*_{1,p}g^*_{1,q}g^*_{1,r} = 0,
\ee
from which one calculates $g^*_{2,i}$. Note that it depends on the yet unknown
$g^{*2}_{1,u}$.
Now $\beta(g^*)=0$ in order $\epsilon^{7/2}$ yields
\be \fl
B_{uu} g^*_{2,u} + C_u g^*_{1,u} =0, \quad
C_u:= 2\sum_p k^0_{upu} g^*_{2,p} +2\sum_p k^1_{upu} g^*_{1,p}
+3\sum_{p,q} l^0_{upqu} g^*_{1,p}g^*_{1,q},
\ee
Since $B_{uu}=0$, we have either $g^*_{1,u}=0$ ({\it RS 2.1a}) or  $C_u=0$,
which
constitutes a quadratic equation in $g^*_{1,u}$ yielding the FP ({\it RS 2.1b}).

Higher orders in $\epsilon$ are determined uniquely. Order $\epsilon^r$, $r>3$
of $\beta_i=0$ yields
\be
\sum_q B_{iq} g^*_{r-1,q} +2k^0_{1uu} g^*_{1,u} g^*_{r-2,u} = {\rm r.h.s},
\ee
where the right hand side of the equation contains $g^*_{r',q}$ with $r'<r-1$
and $g^*_{r',u}$ with $r'<r-2$.
Order $\epsilon^{r+1/2}$ of $\beta_u=0$ yields
\be
B_{uu} g^*_{r-1,u}+C_u g^*_{r-2,u} +2\sum_p k^0_{upu} g^*_{1,u} g^*_{r-1,p} =
{\rm r.h.s}.
\ee
The r.h.s contains $g^*_{r',q}$ with $r'<r-1$ and $g^*_{r',u}$ with $r'<r-2$.
In all cases $B_{uu}=0$. For {\it RS 2.1a}  one
has $g^*_{1,u}=0$ and $C_u\ne 0$, which allows a unique determination of
$g^*_{r-2,u}$. Since each term of the r.h.s. contains at least one factor
$g_{r',u}$, one obtains $g_{r-2,u}=0$. For {\it RS 2.1b} both $B_u=C_u=0$
vanish. However the sum $\sum_p k^0_{upu} g^*_{1,u} g^*_{r-1,p}$ depends via
$g^*_{r-1,p}$ on $g^*_{r-2,u}$. As a result one obtains from this equation
$g^*_{r-2,u}$.

\paragraph{RS 2.1a}
\bea
g^*_{11} = g^*_{22} = \frac 4{n+8}\epsilon
-\frac{4(n^2-2n-20)}{(n+8)^3}\epsilon^2, \nn
g^*_{12} = g^*_{33} = g^*_{13} = g^*_{23} = 0.
\eea
This solution describes two independent \tO{n} models.

\bea
\fl \gamma_{\Phi}^* &=&\left\{ \frac{(n+2)}{4(n+8)^2}\twoeps^2
-\frac{(n+2)(n^2-56n-272)}{16(n+8)^4}\twoeps^3\ (2\times)\right\}, \nn
\fl \gamma_\tau^* &=&\left\{-\frac{n+2}{2(n+8)^2}\twoeps^2, \
-\frac{n+2}{n+8}\twoeps -\frac{(n+2)(13n+44)}{2(n+8)^3}\twoeps^2 \
(2\times) \right\}, \nn
\fl \gamma_{\rm cr}^* &=&\left\{-\frac{2}{n+8}\twoeps
+\frac{(n+4)(n-22)}{2(n+8)^3}\twoeps^2 \ (2\times),
-\frac{n+2}{2(n+8)^2}\twoeps^2 \ (2\times)\right\}, \label{rl-p2-1a} \\
\fl \omega &=&\left\{\twoeps -\frac{3(3n+14)}{(n+8)^2}\twoeps^2 \ (2 \times) ,
\frac{n-4}{n+8}\twoeps +\frac{(n+2)(13n+44)}{(n+8)^3}\twoeps^2, \right. \nn
\fl &&\left.-\frac{n+4}{n+8}\twoeps -\frac{(n+4)(n-22)}{(n+8)^3}\twoeps^2,
\frac{n+2}{2(n+8)^2}\twoeps^2, 0 \right\}. \nonumber
\eea

\paragraph{RS 2.1b}
The second FP to {\it RS 2.1} is given by
\bea
g^*_{11} = g^*_{22} &=& \frac 4{n+8}\epsilon
-\frac{9n^3+98n^2-400n-2272}{2(n+8)^3(n+14)}\epsilon^2, \nn
g^*_{13} = g^*_{23} &=&
\pm\frac{\sqrt{2(n+4)(n+2)(n-4)}}{(n+8)^2\sqrt{n+14}}\epsilon^{3/2}, \\
g^*_{12} &=& - \frac{n+2}{2(n+8)(n+14)}\epsilon^2, \nn
g^*_{33} &=& \frac{(n+2)(n-4)}{(n+8)^2(n+14)}\epsilon^2. \nonumber
\eea
In the limit $D=4$ it is real for $n\ge 4$.
Its critical exponents are
\bea
\fl \gamma_{\Phi}^* &=&\left\{ \frac{(n+2)}{4(n+8)^2}\twoeps^2
\pm \frac{(n+2)\sqrt{2(n-4)(n+2)(n+4)}}{16(n+8)^3\sqrt{n+14}}\twoeps^{5/2}
\right. \nn
\fl && \left. -\frac{(n+2)(n^2-56n-272)}{16(n+8)^4}\twoeps^3\right\}, \nn
\fl \gamma_\tau^* &=&\left\{-\frac{n+2}{(n+8)}\twoeps
-\frac{(n+2)(29n^2+470n+1256)}{4(n+14)(n+8)^3}\twoeps^2, \right. \nn
\fl && \left. -\frac{n+2}{n+8}\twoeps
-\frac{(n+2)(23n^2+434n+1208)}{4(n+8)^3(n+14)}\twoeps^2, \right. \nn
\fl && \left. -\frac{3(n+2)(n^2+10n+64)}{4(n+8)^3(n+14)}\twoeps^2 \right\},
\label{rl-p2-1b} \\
\fl \gamma_{\rm cr}^* &=&\left\{-\frac{2}{n+8}\twoeps
+\frac{n^3-12n^2-660n-2416}{4(n+8)^3(n+14)}\twoeps^2, \right. \nn
\fl && -\frac{2}{n+8}\twoeps
+\frac{3n^3-4n^2-700n-2512}{4(n+8)^3(n+14)}\twoeps^2, \nn
\fl && \left. -\frac{(n+2)(n+6)(n+32)}{4(n+8)^3(n+14)}\twoeps^2,
-\frac{(n+2)(n+26)}{4(n+8)^2(n+14)}\twoeps^2\right\}, \nn
\fl \omega &=&\left\{\twoeps -\frac{3(3n+14)}{(n+8)^2}\twoeps^2 \ (2 \times),
-\frac{n+2}{(n+8)^2}\twoeps^2, 0, \right. \nn
\fl && \frac{n-4}{n+8}\twoeps
+\frac{(n+2)(15n^3+242n^2+656n+32)}{n(n+8)^3(n+14)}\twoeps^2, \nn
\fl && \left.-\frac{n+4}{n+8}\twoeps
-\frac{3n^4+12n^3-332n^2-1252n+64}{n(n+8)^3(n+14)}\twoeps^2
\right\}. \nonumber
\eea

\paragraph{RS 2.3} The representatives of this fixed point are given in
(\ref{rep3}) and (\ref{rep4}).
Its critical exponents are
\bea
\fl \gamma_\Phi^* &=&\left\{
\frac{(2n^6+37n^5+348n^4+2360n^3+9376n^2+13904n-9152)\twoeps^2}{8(n+8)^3(n^2+4n+20)^2}
\right. \nn
\fl && \left.\pm\frac{(n+2)\sqrt{-(3n+22)(n-2)(n+2)(n+4)(n+14)}\twoeps^2}
{8(n+8)^{5/2}(n^2+4n+20)}
\right\}, \nn
\fl \gamma_\tau^* &=&\left\{-\frac{\twoeps(n-1)(n-2)(n+6)}{(n+8)(n^2+4n+20)},
-\frac{(n+2)\twoeps}{2(n+8)} \right. \label{rs-2.3} \\
\fl &&
\left.\pm\frac{\twoeps\sqrt{n^7+32n^6+512n^5+3792n^4+10064n^3-3548n^2-21376n+61184}}{2(n+8)^{3/2}(n^2+4n+20)}
\right\}, \nn
\fl \gamma_{\rm cr}^* &=&\left\{-\frac{\twoeps}{n+8} \pm
\frac{\sqrt{-2(n^5+34n^4+312n^3+752n^2-1776n-7648)}\twoeps}{(n+8)^{3/2}(n^2+4n+20)},
\right. \nn
\fl && \left. -\frac{(n+6)(3n+2)\twoeps} {(n+8)(n^2+4n+20)},
-\frac{(n+6)(n+14)\twoeps}{(n+8)(n^2+4n+20)}
\right\}, \nn
\fl \omega &=&\left\{0, \twoeps,
\frac{\twoeps(n^3+10n^2-4n-232)}{(n+8)(n^2+4n+20)} ,
\frac{\twoeps\lambda'}{2(n+8)(n^2+4n+20)} \right\}. \nonumber
\eea
where $\lambda'$ is solution of the equation
\bea
\fl && \lambda^{\prime 3}+16(n^2+4n+20)\lambda^{\prime 2}
-4(n+4)(n^5-18n^4-392n^3-1648n^2-496n+8928)\lambda' \nn
\fl && -16(3n+22)(n-2)(n+6)(n-6)(n+4)(n+2)(n+14)^2=0.
\eea
In an expansion in $1/(n+8)$ one obtains the $\omega$s
\bea
\twoeps && \left(-\frac{6}{n+8} -\frac{296}{(n+8)^2} -\frac{11272}{(n+8)^3}
+O(\frac 1{(n+8)^4})\right) \nn
\twoeps && \left(1-\frac{20}{n+8} -\frac{78}{(n+8)^2} +\frac{906}{(n+8)^3}
+O(\frac 1{(n+8)^4})\right), \\
\twoeps && \left(-1+\frac{18}{n+8}+\frac{374}{(n+8)^2}+\frac{10366}{(n+8)^3}
+O(\frac 1{(n+8)^4})\right). \nonumber
\eea

Similarly as for {\it RS 1.4} we consider the coupling in two loop order, since
it yields in order $\epsilon$ the region in which the couplings are real. Using
the representation (\ref{rep3}) the couplings may be written in the form
(\ref{U1}, \ref{V1}) with
\bea
\fl u^{(1)} &=& \frac 2{n+8}, \nn
\fl u^{(2)} &=& -\frac{2(n^7 +21n^6 +249n^5+ 1564n^4 +2312n^3 -13808n^2
-53104n-55360)}{(n^2 +4n + 20)^2 (n+8)^4}, \nn
\fl v^{(1)} &=& \frac 2{(n+8)^2(n^2+4n+20)}, \nn
\fl v^{(2)} &=& \frac{-2(n+2)(n+4)
\left({3n^{11} +13n^{10} -2301n^9 -41840n^8 -134712n^7
 +2573392n^6 +26618112n^5 \atop +82530752n^4 -6879104n^3 -368123392n^2
 -274477824n -126516224}\right)}{(n+14)(n+8)^4(n^2+4n+20)^3}, \nn
\fl w &=& -(3n+22)(n-2)(n+2)(n+4)(n+8)(n+14), \\
\fl g^*_{12} &=& \frac{4(n+4)(n+6)}{(n+8)(n^2+4n+20)}\epsilon
+\frac{2\left({n^{10} -21n^9 -1596n^8 -24396n^7 -124064n^6 +251792n^5 \atop
 +5029824n^4 +19095232n^3 +27139840n^2 +6788096n -8542208}\right)}
{(n+14)(n+8)^4(n^2+4n+20)^3} \epsilon^2, \nn
\fl g^*_{33} &=& \frac{4(n^2-36)}{(n+8)(n^2+4n+20)} \epsilon
- \frac{4\left({n^{10} +30n^9 -99n^8 -13222n^7 -189636n^6 -1087512n^5 \atop
 -1638768n^4 +8148960n^3 +31543872n^2 +18656640n
-30614016}\right)}{(n+14)(n+8)^4(n^2+4n+20)^3} \epsilon^2, \nn
\fl g^*_{13} &=& g^*_{23} = 0. \nonumber
\eea
From (\ref{V2}, \ref{V3}) we obtain
$n_{\rm c}=2-\twoeps/140+O\twoeps^2$.

\paragraph{RS 2.2 and 3.1} These two fixed points are solutions of one and the
same quadratic equation. Both fixed points are \tO{n}$\times$\tO{2} invariant.
In two-loop order the solutions $g^*$ can be written
\bea
g_{ij}^* &=& u^{(1)}_{ij}\epsilon + u^{(2)}_{ij}\epsilon^2 +sV_{ij}, \nn
V_{ij} &=& v^{(1)}_{ij} \sqrt w \epsilon + \frac{v^{(2)}_{ij}}{\sqrt w}
\epsilon^2, \nn
g_{11}^* &=& g_{22}^*, \quad g_{13}^*=g_{23}^* = 0, \nn
u_{11}^{(1)} &=&\frac{3n^2-2n+24}N, \quad v_{11}^{(1)} =\frac{n-6}N,
\label{FP-10} \\
u_{12}^{(1)} &=&-\frac{(n+12)(n-6)}N, \quad v_{12}^{(1)} =\frac{n+6}N, \nn
u_{33}^{(1)} &=&\frac{4(n-3)(n-4)}N, \quad v_{33}^{(1)} =\frac{-12}N, \nn
N &=& n^3+4n^2-24n+144, \nn
w &=& n^2-24n+48, \nonumber
\eea
where $s=+1$ corresponds to {\it RS 3.1} called chiral FP, and $s=-1$ to {\it RS
2.2} is called antichiral.
Close to $D=4$ they are real only for $n\ge22$ and $n\le2$.
The critical exponents read
\bea
\fl \gamma_{\Phi}^* &=&
\left\{\frac{(5n^5-3n^4-16n^3-656n^2+3072n-1152+s(n-3)(n+4)w^{3/2})\twoeps^2}{16N^2}
\ (2\times) \right\}, \label{Tau-10} \nn
\fl \gamma_{\tau}^* &=&\left\{-\frac{(n
(48+n+n^2)+s(n-3)(4+n)\sqrt w) \twoeps}{2N}, \right. \\
\fl && \left. \frac{(-2n^3-3n^2+28n-48
+5s n \sqrt w)\twoeps }{2N} \ (2\times)\right\}.
\nonumber
\eea
The exponent $\gamma_{\tau_1}^*$  determines $\nu$, whereas the two degenerate
ones yield cross-over exponents.

The other cross-over exponents are obtained from
\bea
\fl \gamma_{\rm cr}^* &=& \left\{
\frac{(-5n^2-s(n-12)\sqrt w)\twoeps}{2N}, \
\frac{(-n^2+4n-48-sn\sqrt w)\twoeps}{2N} \
(2\times), \right.
\label{FP-10-gammacr} \\
\fl && \left.
\frac{(3n^2+8n-96-s(n+12)\sqrt w)\twoeps}{2N}
\right\}. \nonumber
\eea
The six exponents $\omega$ are
\bea
\fl \omega &=&\left\{
\frac{(n+4)\big((n+4)(n-3)-3s\sqrt w \big)\twoeps}
N \ (2\times), \right.
\label{FP-10-omega} \\
\fl && \frac{(n^3+14n^2+56n-96+s(n+8)(n-6)\sqrt w )\twoeps}
{2N} \ (2\times),
\nn
\fl && \left. \frac{(-3(n^2-24n+48)+s(n+4)(n-3)\sqrt w )\twoeps}N,
 \ \twoeps \right\}.
\eea
In two-loop order one can write
\bea
u^{(2)}_{11} &=& \frac 1{N^3} \left({-3n^8 +10n^7 -432n^6 +1710n^5 +7480n^4
\atop +20976n^3 +3456n^2 -411264n +456192}\right), \nn
u^{(2)}_{12} &=& \frac 1{N^3} \left({n^8 +20n^7 -286n^6 -3550n^5 -11960n^4 \atop
+32208n^3 +165888n^2 +148608n -290304}\right), \nn
u^{(2)}_{33} &=& -\frac{2(n-3)(n+4)}{N^3} \left({2n^6 +3n^5 +94n^4 -2688n^3
\atop -5904n^2 -20736n +31104}\right), \nn
v^{(2)}_{ij} &=& \frac{F_{ij}w}{N^3} + \frac{v^{(1)}_{ij}L}{N^2}, \\
F_{11} &=& -n^7 +50n^6 +552n^5 +11726n^4 +230912n^3 +5022864n^2 \nn
&& +109907904n -910069632, \nn
F_{12} &=& -n^7 +4n^6 +658n^5 +19546n^4 +416192n^3 +8896560n^2 \nn
&& +193441728n +909986688, \nn
F_{33} &=& 46n^6 -106n^5 -7820n^4 -185280n^3 -3873696n^2 \nn
&& -83533824n -1820056320, \nn
L &=& 11520000 (287n-632). \nonumber
\eea
As a result one obtains
\be
n_{\rm c} = n_0 -\frac{2L(n_0)}{N^2(n_0)w'(n_0)}\epsilon
\ee
which yields $n_c=12\pm 4\sqrt 6-(12\pm14\sqrt 6/3)\twoeps+O\twoeps^2$ in
agreement with \cite{Kawamura-1988}.\footnote{Eq. (4.5) in \cite{Vicary-622} is
misprinted. The correct result is found in (3.10) of \cite{Kawamura-1988}.}

The FP (\ref{FP-10}) is stable for large $n$, where the sign $s$ in front of the
root $\sqrt{w}$ is chosen positive. The stability of this FP in three
dimensions is discussed on the basis of various calculation schemes in sect.
11.5.3 of \cite{Review}, see also
\citef{16-21,\,30-32}{Antonenko1,Antonenko,Loison,Pelissetto1,Calabrese,Calabrese1,%
Delamotte1,Loison1,Thanh}.

The large $n$-expansion of critical exponents for the FP (\ref{FP-10}) was
performed in
\cite{Kawamura-1988,Weber}. We mention the results for the
exponents $\gamma_\Phi^*$ and $\gamma_\tau^*$ in arbitrary dimension
$D$ and in the first order of $1/n$
\bea
\eta = \frac{6 \Gamma(D-2) \sin(\frac{D\pi}{2})}{\pi \Gamma(D/2 -2)
\Gamma(1 + D/2)n}
\\
\fl 1/\nu=D-2 + \frac{2(2-D)(1-D)\eta}{4-D}, \quad
1/\nu_2=D-2 + \frac{2(2-D)(3-2D)\eta}{3(4-D)}. \label{1_n}
\eea
The exponents $\eta$ and $\nu$ were already given in \cite{Kawamura-1988}, the exponent $\nu_2$ in \cite{Weber},
where $1/\nu_2 = 2+\gamma^*_{\tau,2}$. It yields a cross-over exponent
$\Delta=\nu/\nu_2$, compare eqs. (\ref{Fsing},\ref{gapexp}).

\section{The well-known Subcases\label{subcases}}

Here we review the FPs of our model (\ref{E-1}) which also contain the
actions (\ref{E-0a}-\ref{E-0c}). While the trivial
Gaussian FP is unstable in all models, their stable FPs, apart of
the stable FP of (\ref{E-0c}), are found unstable in the
general model (\ref{E-1}).

The  nontrivial $n$-Heisenberg FP of the simple
$\phi^4$ model (\ref{E-0a}) is stable and corresponds to
{\it RS 1.1}, if the second field is neglected. The quantities
(\ref{rl-p1-1}) reduce to
\be \fl
\gamma_\Phi^*=\frac{(n+2)\twoeps^2}{4(n+8)^2}, \quad
\gamma_\tau^*=-\frac{(n+2)\twoeps}{n+8}, \quad
\gamma_{\rm cr}^* = -\frac{2\twoeps}{n+8}, \quad
\omega=\twoeps. \label{phi4-exp}
\ee

The models (\ref{E-0b}) and (\ref{E-0c}) are special cases of model (\ref{E-1}).
Since the number of independent couplings $\tau$ and $g$ are less, the number of
exponents $(\gamma_{\Phi}^*,\gamma_{\tau}^*,\omega)$ reduce to (2,2,3) for
model (\ref{E-0b}) and to (1,1,2) for model (\ref{E-0c}). Those exponents
$\gamma_{\tau}^*$ of (\ref{E-1}), which are no longer $\gamma_{\tau}^*$s of
(\ref{E-0b}) and (\ref{E-0c}) belong now to the exponents $\gamma_{\rm cr}^*$.

The \tO{n}+\tO{n} model (\ref{E-0b}) with
$3g_{11}=g_1, 3g_{22}=g_2, 6g_{12}=g_3, g_{13}=0, g_{23}=0, g_{33}=0$
has six nontrivial FPs.
Three of them are decoupled ($g_3^*=0$) and therefore represent
tetracritical rather than bicritical behavior \cite{Fisher}: The
$n$-Heisenberg-Gaussian FP with $g_1^*=6\twoeps/(n+8)$
and  $g_2^*=0$, the Gaussian-$n$-Heisenberg FP with $g_2^*=6\twoeps
/(n+8)$ and  $g_1^*=0$ and the
$n$-Heisenberg-$n$-Heisenberg FP with $g_1^*=g_2^*=6\twoeps/(n+8)$.
The critical exponents of the
$n$-Heisenberg-Gaussian FP {\it RS 1.1} in (\ref{rep3}) are
\bea
\gamma_\Phi^* &=& \left\{\frac{(n+2)\twoeps^2}{4(n+8)^2},0\right\},\quad
\gamma_\tau^*=\left\{-\frac{(n+2)\twoeps}{n+8}, 0\right\}, \nn
\omega &=& \left\{\twoeps,-\twoeps,-\frac{6\twoeps}{n+8}\right\},
\eea
and the  decoupled $n$-Heisenberg-$n$-Heisenberg FP {\it RS 2.1} in (\ref{rep1})
has
\bea
\gamma_{\Phi}^* &=& \left\{\frac{(n+2)\twoeps^2}{4(n+8)^2}\ (2\times)\right\},
\quad
\gamma_{\tau}^*=\left\{-\frac{(n+2)\twoeps}{n+8}\ (2\times)\right\}, \nn
\omega &=& \left\{\twoeps\ (2\times) , \frac{(n-4)\twoeps}{n+8}\right\}.
\eea
The latter FP is clearly  stable for $n>4$.

The three remaining FPs have a nonvanishing $g_3^*$ and therefore
represent bicritical behavior. The first FP is the isotropic
$2n$-Heisenberg FP \cite{Wegner-Fi, Fisher-Pfeuty} {\it RS 1.3} in (\ref{rep1})
with
\bea
\gamma_{\Phi}^* &=&
\left\{\frac{(2n+2)\twoeps^2}{4(2n+8)^2}\right\}, \quad
\gamma_{\tau}^*=\left\{-\frac{(2n+2)\twoeps}{2n+8},
-\frac{2\twoeps}{2n+8}\right\}, \nn
\omega &=& \left\{\twoeps,\frac{8\twoeps}{2n+8},
-\frac{(2n-4)\twoeps}{2n+8}\right\}.
\eea
This FP is stable for $n<2$.
The first $\gamma_{\tau}^*\sim O(\epsilon)$ is the true critical
exponent for $\tau$, the second $\gamma_{\tau}^*\sim O(\epsilon/n)$ yields the
cross-over exponent.

The second FP is the so called biconical FP {\it RS 1.2} in (\ref{rep1}). Its
critical exponents are
\bea
\gamma_{\Phi}^* &=&\left\{\frac{n(n^2-3n+8)\twoeps^2}{8(n^2+8)^2}\
(2\times)\right\}, \quad
\gamma_{\tau}^*= \left\{\frac{(1-n) n\twoeps}{n^2+8},
-\frac{3n \twoeps}{n^2+8}\right\},
\nn
\omega &=&\left\{\twoeps,\frac{8(n-1)\twoeps}{8+n^2},
\frac{(4-n) (n-2) \twoeps}{8+n^2}\right\}.
\eea
The biconical FP is stable for $n=3$ in our approximation.

The last FP is given by {\it RS 1.4} in (\ref{rep3}) and is
complex for $n>1$. Its critical exponents are given by $\gamma_{\phi1,2}^*$,
$\gamma_{\tau1,2}^*$, and $\omega_{2,5,6}$ of (\ref{rs14}).
This FP coincides with the  biconical FP for $n=1$.

The frustrated spin model (\ref{E-0c}) is invariant under \tO{n}$\times$\tO{2}.
It is obtained by
$\tau_2=\tau_1$, $g_{11}=g_{22}=u/3$, $g_{12}=u/3-v/6$, $g_{33}=v/6$,
$g_{13}=g_{23}=0$. It has four FPs: the trivial
Gaussian FP {\it RS 0.1}, the isotropic $2n$-Heisenberg FP {\it RS 1.3},
and the fixed point {\it RS 2.2} and {\it RS 3.1}. $\gamma_{\Phi}^*$ is that of
{\it RS 2.2} and {\it 3.1}. $\gamma_{\tau}$ is $\gamma_{\tau_1}^*$ of
(\ref{Tau-10}), the other $\gamma_{\tau_i}$ and the $\gamma_{\rm cr}$ of
(\ref{Tau-10}) yield the cross-over exponents, and $\omega$ equals
$\omega_{5,6}$ of (\ref{FP-10-omega}).

The FPs 2.2 and 3.1 are complex for $2.2<n<21.8$ close to $D=4$. This region
decreases with decreasing $D$\cite{Vicary-622,Antonenko}. The question of the
range of stability in $D=3$ is under debate%
\citef{4,\,16-21,\,30-32}{Review,Antonenko1,Antonenko,Loison,Pelissetto1,Calabrese,%
Calabrese1,Delamotte1,Loison1,Thanh}.

\section{Summary and Conclusion}

We considered in detail the \tO{n}-model (\ref{E-1}) of two fields.

We gave the expressions for the $\beta$ functions
(\ref{WV1},\ref{beta_coeff}) and the matrices $\gamma_\Phi$
(\ref{the-4-eps-gamma-phi}), $\gamma_\tau$
(\ref{the-4-eps-gamma-tau}), $\gamma_{\rm cr,s}$ (\ref{gamma_crs}) and $\omega$
(\ref{the-4-eps-omega}), and $\gamma_{\rm cr,a}$ (\ref{gamma_cra}) for the model
(\ref{E-1}) from which the critical exponents are obtained in one-loop order
(for $\eta$ in two-loop order).

Next we considered its
properties under orthogonal transformations of the two fields.
Two types of FPs emerge: Four of them are invariant under \tO{n}$\times$\tO{2}.
The other FPs are not invariant under \tO{2} and yield lines of FPs. The
transformation of the couplings under \tO{2} were given.

A classification of the FPs in the large $n$-limit was given, before they were
determined for general $n$. Under the
numerous FPs the corresponding FPs of the well-known models were
found. To our best knowledge the FPs {\it RS 2.1b} and {\it 2.3} are new.
{\it RS 2.1b} has the remarkable property that it agrees for arbitrary $n$ with
{\it RS 2.1a}, which describes two uncoupled systems, in one-loop order.
For these FPs two of the exponents $\omega$ vanish in one-loop order. For
special values of $n$ some of the FPs coincide or yield an extra vanishing
$\omega$. This is left for further discussion.

The full description of the fixed-point structure
and the values of the most essential critical exponents can be useful for
analytical and numerical investigation of the critical features of
the system near $D=4$. In this way a better understanding of such
interesting
phenomena as inverse symmetry breaking, symmetry nonrestoration, and
reentrant phase transitions could be obtained. Our model generalizes the
\tO{n}+\tO{n} and the \tO{n}$\times$\tO{2} model giving rise to a variety of
multi-critical phenomena.

\section*{Acknowledgements}

We are grateful to A.I. Sokolov, Yu. Holovatch and D. Mouhanna for interest in our paper, fruitful discussions and useful suggestions. Yu.M. Pis'mak was supported in part by the Russian Foundation of Basic Research (RFRB grant $07$--$01$--$00692$).
A. Weber has been supported by a grant of the LGFG Baden-W\"urttemberg.

\begin{appendix}

\section{Stability matrix in the large $n$ limit\label{stabmat}}

The stability matrix in one-loop order is given by
\be
\fl \omega=\frac{\partial\beta}{\partial g} =
-2\epsilon 1_6 +\frac{n+8}2
\left(\begin{array}{cccccc} 2g_{11} & 2 g_{12} & 2 g_{13} & 0 & 0 & 0 \\
g_{12} & g_{11}+g_{22} & g_{23} & g_{12} & g_{13} & 0 \\
g_{13} & g_{23} & g_{11}+g_{33} & 0 & g_{12} & g_{13} \\
0 & 2g_{12} & 0 & 2g_{22} & 2g_{23} & 0 \\
0 & g_{13} & g_{12} & g_{23} & g_{22}+g_{33} & g_{23} \\
0 & 0 & 2g_{13} & 0 & 2g_{23} & 2g_{33}
\end{array}\right)
\ee
and yields in the large $n$ limit for $p^{(1)}$
\be
\omega = -2\epsilon 1_6 + 2\epsilon B_1 B_2
\ee
with
\be
B_1=\left(\begin{array}{ccc} 2\gam_1 & 0 & 0 \\ \gam_2 & \gam_1 & 0 \\
\gam_3 & 0 & \gam_1 \\ 0 & 2\gam_2 & 0 \\
0 & \gam_3 & \gam_2 \\ 0 & 0 & 2\gam_3 \end{array}\right),
\quad
B_2=\left(\begin{array}{cccccc}
\gam_1 & \gam_2 & \gam_3 & 0 & 0 & 0 \\
0 & \gam_1 & 0 & \gam_2 & \gam_3 & 0 \\
0 & 0 & \gam_1 & 0 & \gam_2 & \gam_3 \end{array}\right).
\ee
The $3\times3$-matrix $B_2B_1$
\be
(B_2B_1)_{ij} = \delta_{ij}+\gam_i\gam_j
\ee
has one eigenvalue 2 and two eigenvalues 1. The matrix $B_1B_2$ has the same
eigenvalues and in addition three eigenvalues 0. As a consequence the stability
matrix $\omega$ has three eigenvalues $-2\epsilon$, two eigenvalues 0 and one
eigenvalue $2\epsilon$ independent of $\gam_{12}$.

For $p^{(2)}$ the stability matrix reads in this limit
\be
\omega = 2\epsilon 1_6 - 2\epsilon B_1 B_2
\ee
and thus the eigenvalues are the negative of those of $\omega$ for
$p^{(1)}$.

\end{appendix}

\end{document}